\documentclass[twocolumn]{aastex631} 

\begin{document}
\title{Primordial Black Hole Merger Rate in $f(R)$ Gravity}

\author[0000-0002-6349-8489]{Saeed Fakhry} 
\email{s\_fakhry@sbu.ac.ir}
\affiliation{Department of Physics, Shahid Beheshti University, Evin, Tehran 19839, Iran}

\begin{abstract}
Primordial black holes (PBHs) are known as one of the potential candidates for dark matter. They are expected to have formed due to the direct gravitational collapse of density fluctuations in the early Universe. Therefore, the study of the merger rate of PBHs in modified theories of gravity can provide more detailed information about their abundance. In this work, we delve into the calculation of the merger rate of PBHs within the theoretical framework of $f(R)$ gravity. Our analysis reveals an enhancement in the merger rate of PBHs compared to that obtained from general relativity (GR). Additionally, modulating the field strength $f_{\rm R0}$ induces shifts in the PBH merger rate, presenting a potential observational signature of modified gravity. We also find that the total merger rate of PBHs will be consistent with the merger rate of black holes estimated by the Laser Interferometer Gravitational-Wave Observatory (LIGO)-Virgo-KAGRA detectors if $f_{\rm PBH}\gtrsim 0.1$. While further improvements might be required, relative enhancement of the merger rate of PBHs in the framework of $f(R)$ gravity and its consistency with gravitational wave data underscore the importance of employing modified theories of gravity to examine diverse scenarios related to the formation of black holes.
\end{abstract}


\section{Introduction} 
Gravitational waves (GWs), as cosmological phenomena, have been a central area of investigation for several decades. Consequently, the detection of GWs has introduced a novel framework for analyzing astrophysical and cosmological phenomena. One of the main sources of GWs is the merging of binary black holes, which produces distinguishable signals in GW detectors \citep{2022LRR....25....1M, 2022PhRvD.106d3009O}. In recent years, numerous binary black hole mergers have been identified through the collaborative efforts of the Laser Interferometer Gravitational-Wave Observatory (LIGO)-Virgo-KAGRA detectors \citep{2016PhRvL.116f1102A, 2016PhRvL.116x1103A, 2016PhRvL.116v1101A, 2020ApJ...896L..44A, 2020PhRvL.125j1102A}. This accomplishment has opened a fresh avenue for comprehending compact objects distributed throughout the Universe.

Notably, the binary black hole mergers identified by the LIGO-Virgo-KAGRA observatories are related to those of stellar mass, i.e., those with a mass less than $100\,M_{\odot}$. This revelation holds the potential to offer significant insights into the distribution of black hole masses across the Universe, as well as their mechanisms of formation. The exact processes responsible for the creation of black holes involved in binary merger events detected by the LIGO-Virgo-KAGRA detectors are yet to be conclusively determined. These black holes could originate from the collapse of massive stars (potentially through diverse channels) \citep{2021ApJ...912...98F, 2021RNAAS...5...19R}, or they might have arisen from the direct collapse of primordial density fluctuations during the early stages of the Universe \citep{2016PhRvL.116t1301B, 2016PhRvL.117f1101S, 2017PDU....15..142C}.

It is captivating to note that the GW information gathered by the LIGO-Virgo-KAGRA detectors corresponds closely to the idea of mergers involving PBHs, which are among the potential explanations put forth to account for dark matter, see, e.g., \citep{2000PhRvL..84.3760S, 2000ApJ...542..281A, 2012LRR....15....6L, 2018RPPh...81f6201R, 2020PhRvL.124j1303B, 2020PhRvD.102h4063D, 2021CQGra..38s4001D, 2023arXiv230712924S}. The formation of PBHs is theorized to occur due to nonlinear peaks within initial density fluctuations. In fact, when these fluctuations surpass a certain threshold value, they may directly collapse into PBHs \citep{1967SvA....10..602Z, 1971MNRAS.152...75H, 2012RPPh...75l4901V}. Various studies have explored the development of nonlinear density fluctuations on scales larger than the horizon, as well as the critical amplitude of curvature perturbations required for black hole formation \citep{1999PhRvD..59l4013N, 1999PhRvD..60h4002S, 2007CQGra..24.1405P, 2014JCAP...07..045Y, 2017JCAP...06..041A}. Additionally, PBHs can cover a broad spectrum of masses, setting them apart from astrophysical black holes \citep{2018CQGra..35f3001S}.

Due to their stochastic spatial distribution, PBHs within the confines of dark matter halos could potentially encounter each other or other compact objects, forming binary systems that emit GWs and ultimately undergo merger. In recent years, the detection of GWs stemming from the coalescence of binary black holes by the LIGO-Virgo-KAGRA detectors has sparked renewed fascination with PBHs \citep{2019PhRvX...9c1040A, 2021PhRvX..11b1053A, 2021arXiv211103606T}. Given the mass range of black holes engaged in binary mergers, which often surpasses the spectrum of masses observed in typical astrophysical black holes, it becomes plausible that these black holes originated in early Universe. Considering PBHs as conceivable candidates for dark matter and acknowledging the possibility of their clustering within dark matter halos, the structural properties of these halos could potentially influence the merger rate of PBHs. Thus, it has been suggested that certain quantities, such as the halo mass function, halo concentration parameter, and halo density profile, could enhance the precision of theoretical models and their prognostications concerning the merger rate of PBHs. For more details, see our previous works \citep{2021PhRvD.103l3014F, 2022PhRvD.105d3525F, 2022ApJ...941...36F, 2023PhRvD.107f3507F, 2023ApJ...947...46F, 2023PDU....4101244F}. 

Several models have been suggested to offer an elaborate depiction of the halo mass function \citep{1974ApJ...187..425P, 1999MNRAS.308..119S, 2001MNRAS.323....1S, 2001MNRAS.321..372J, 2003MNRAS.346..565R, 2006ApJ...646..881W, 2007MNRAS.374....2R, 2017JCAP...03..032D}, and halo concentration parameter \citep{2012MNRAS.423.3018P, 2016MNRAS.456.3068O, 2016MNRAS.460.1214L, 2017arXiv171105277O} within the context of GR. However, a fundamental inquiry arises regarding whether the principles of GR and the ideas derived from it provide an adequate framework for explaining the formation and evolution of large-scale structures like galactic halos or not. Nowadays, there is a growing consensus that, while GR has achieved impressive feats in accurately predicting numerous events, it falls short when it comes to explaining large-scale phenomena such as dark sectors of the Universe, see, e.g., \citep{2011PhRvL.107b1302S, 2014JCAP...08..034Y, 2018RPPh...81a6901H, 2020PDU....3000666D, 2020IJMPD..2930014A, 2021arXiv210700562G, 2022arXiv221111273B, 2023PDU....3901144G}. Therefore, to precisely characterize these dark sectors of the universe, numerous endeavors have been carried out to extend the Einstein-Hilbert action, resulting in what are referred to as modified theories of gravity.

One of the renowned modified theories of gravity is known as $f(R)$ gravity, wherein a function of the Ricci scalar is substituted for the Ricci scalar itself within the Einstein-Hilbert action. To some extent, $f(R)$ gravity has demonstrated the ability to address the characterization of dark sectors of the Universe. However, some studies have shown that general $f(R)$ gravity faces challenges in simultaneously satisfying cosmological constraints and solar system tests of gravity \citep{2011IJMPD..20.1357G, 2014IJMPD..2350036G, 2020PhRvD.101f4005N}. In this context, it has been demonstrated that certain categories of $f(R)$ gravity theories offer a viable solution to this matter to a satisfactory degree. On of these models, known as the Hu-Sawicki model \citep{2007PhRvD..76f4004H}, introduces a scalar field, known as the scalaron, which is derived from the modification of gravitational theory. An interesting property of the scalaron in this model is its chameleonic nature, whereby it becomes light in low-density regions and heavy in high-density regions of matter. The Hu-Sawicki model is one of the viable models within the framework of $f(R)$ gravity and has been used to explain dark sectors of the Universe with higher accuracy, see, e.g., \citep{2009PhRvD..79l3516M} and references therein.

As a result, enhancing the precision of theoretical forecasts concerning the dark sectors of the Universe holds the promise of yielding a more comprehensive picture of the abundance of PBHs, which are being considered as potential candidates for dark matter. Consequently, our aim is to compute the merger rate of PBHs within the framework of $f(R)$ gravity. In this respect, the outline of the work is as follows: In Sec.\,\ref{sec:ii}, we describe the theoretical framework of $f(R)$ gravity. In Sec.\,\ref{sec:iii}, we elucidate the concepts of dark matter halo models and essential factors, such as the halo density profile, the halo concentration parameter, and the halo mass function, within the theoretical contexts of both GR and $f(R)$ gravity. Moreover, in Sec.\,\ref{sec:iv}, we specify the merger rate of PBHs in $f(R)$ gravity and compare it with the corresponding findings from GR. We also constrain the abundance of PBHs in the framework of $f(R)$ gravity while considering various masses of PBHs. Finally, in Sec.\,\ref{sec:v}, we discuss the results and summarize the findings.
\section{Theoretical Framework of $f(R)$ Gravity} \label{sec:ii}
In this section, we intend to explain the theoretical framework of $f(R)$ gravity. In Jordan frame, the action of $f(R)$ gravity is defined by \citep{2012PhRvD..85l4054L}
\begin{equation}
S=\frac{1}{2\kappa}\int d^{4}x\sqrt{-g}[R + f(R)] + S_{m}(g_{\mu \nu}, \psi_{m}),
\end{equation}
in which $R$ stands for the Ricci scalar, $\kappa$ represents the Einstein gravitational constant, $g$ is the determinant of the metric tensor $g_{\mu\nu}$, and $S_{m}$ denotes the action associated with matter fields $\psi_{m}$. It is important to observe that by selecting $f = -2\Lambda$, GR can be reinstated along with a cosmological constant. It is worth emphasizing that different models can be discerned by the specific $f(R)$ function chosen. Modifying the metric through variation yields the adjusted field equations for $f(R)$ gravity.
\begin{equation}
G_{\mu\nu}+f_{R}R_{\mu \nu} -\frac{1}{2}f(R)g_{\mu \nu} -\nabla_{\mu}\nabla_{\nu}f_{R}+g_{\mu\nu}\square f_{R}=\kappa T_{\mu\nu},
\end{equation}
where $G_{\mu\nu}=R_{\mu\nu}-1/2R g_{\mu\nu}$ is defined as the Einstein tensor, and $R_{\mu\nu}$ represents the Ricci curvature tensor. The additional scalar degree of freedom, termed the scalaron and denoted by $f_{R} \equiv df(R)/dR$, emerges from the theory and characterizes the alterations in forces. $T_{\mu\nu}$ stands for the energy-momentum tensor. The Levi-Civita-like connection is denoted as $\nabla_{\mu}$, and the d'Alembert operator is abbreviated as $\square \equiv \nabla_{\alpha}\nabla^{\alpha}$. It is evident that the aforementioned field equations involve derivatives up to the fourth order. Consequently, these equations can be interpreted as the standard field equations for GR supplemented by an additional scalar field $f_{R}$. Furthermore, taking the trace of the modified field equations permits the derivation of the equation of motion for the scalaron
\begin{equation} \label{scalaron_fe}
3\square f_{R}+f_{R}R-2f(R)=\kappa \rho_{\rm m},
\end{equation}
where $\rho_{\rm m}$ is the matter density of the Universe. Given the quasistatic approximation for a flat Universe and scales sufficiently below the horizon, one can neglect the time derivative of $f_{R}$ in the field equations. Hence, Eq. (\ref{scalaron_fe}) can be simplified as
\begin{equation}
\frac{1}{a^{2}}\bigtriangledown^{2}f_{R}\approx-\frac{1}{3}[R-\bar{R}+\kappa (\rho_{\rm m}-\bar{\rho}_{\rm m})],
\end{equation}
in which $\rho_{\rm m}$ denotes the density of matter in the Universe. Assuming a quasistatic approximation for a flat Universe and focusing on scales significantly smaller than the horizon, one can disregard the temporal change of $f_{R}$ within the field equations. As a result, Eq.\,(\ref{scalaron_fe}) can be rendered more straightforward as follows:
\begin{equation}
\frac{1}{a^{2}}\bigtriangledown^{2}\Phi\approx\frac{1}{6}(R-\bar{R})+\frac{\kappa}{3}(\rho_{\rm m}-\bar{\rho}_{\rm m}),
\end{equation}
which addresses the Newtonian potential $\Phi$ within the framework of $f(R)$ gravity. Ensuring highly precise forecasts akin to those of GR within the solar system holds substantial importance, particularly when adjusting the gravitational theory to account for the observed accelerated expansion of the Universe on large scales.

On the other hand, $f(R)$ gravity within Jordan frame can be converted into its counterpart in Einstein frame through a conformal transformation applied to the metric \citep{2012PhRvD..85l4054L}
\begin{eqnarray} 
S_{\rm E} =\int d^{4}x\sqrt{-\tilde{g}}\left[\frac{\tilde{R}}{2\kappa}-\frac{1}{2}\tilde{\partial}^{\mu}\phi\tilde{\partial}_{\mu}\phi-V(\phi)\right] \nonumber \\
 +S_{m}[A^{2}(\phi)\tilde{g}_{\mu\nu}, \psi_{m}],
\end{eqnarray}
where $A^{2}(\phi)$ is the conformal factor and a tilde demonstrates quantities in Einstein frame, and
\begin{equation}
\tilde{g}_{\mu\nu} \equiv (1+f_{R})g_{\mu\nu}, 
\end{equation}
\begin{equation} \label{phi}
\left(\frac{d\phi}{df_{R}}\right)^{2} \equiv \frac{3}{2\kappa(1+f_{R})^{2}},
\end{equation}
\begin{equation}
A(\phi) \equiv \frac{1}{\sqrt{1+f_{R}}},
\end{equation}
\begin{equation}\label{potentialphi}
V(\phi) \equiv \frac{f_{R} R - f(R)}{2\kappa (1+f_{R})^{2}}.
\end{equation}
The integration of Eq.\,(\ref{phi}) yields the subsequent formulation for the scalar field
\begin{equation}
\phi = \sqrt{\frac{3}{2\kappa}}\ln(1+f_{R})+\phi_{0}.
\end{equation}
Variation of the action with respect to $\phi$ results in:
\begin{equation}\label{tildephieq}
\tilde{\square}\phi = - \alpha \tilde{T} +V^{\prime}(\phi) \equiv V^{\prime}_{\rm eff}(\phi),
\end{equation}
where $\alpha = d \ln A/d\phi$ and $V_{\rm eff}$ represents the effective potential dictating the behavior of $\phi$. It is worth mentioning that $\tilde{T} = A^{4}(\phi)T$. In the quasistatic regime, one can disregard temporal derivatives in Eq.\,(\ref{tildephieq}), leading to the scalar field equation
\begin{equation}
\nabla^{2}\phi = \alpha A^{4}(\phi)\rho_{\rm m} + V^{\prime}(\phi),
\end{equation}
which is written in physical coordinate. When selecting a specific functional form for $f(R)$, it is important to consider that the chosen model should be able to effectively explain observed phenomena. Therefore, an appropriate $f(R)$ model should adhere to the following criteria: {\it (i)} The cosmological behavior emerging from the proposed model should closely resemble that of the standard model in the early Universe, a stage thoroughly examined through Cosmic Microwave Background (CMB) testing. {\it (ii)} The model must facilitate the acceleration of the Universe's expansion in the late-time phase, without relying on a cosmological constant. This acceleration should closely align with the expansion history predicted by the standard model of cosmology. {\it (iii)} The chosen model should possess adequate degrees of freedom to encompass a wide array of observed phenomena in the late-time Universe. {\it (iv)} The model should incorporate the standard model's phenomenology as a limiting scenario, enabling the ability to constrain minor deviations from GR in assessments involving both large and small scales. As a result, the validity of model hinges on satisfying these aforementioned conditions
\begin{eqnarray}
\lim_{R\rightarrow\infty} f(R) = \rm const.,\nonumber \\
\lim_{R\rightarrow 0} f(R) = 0.
\end{eqnarray}
In \citep{2007PhRvD..76f4004H}, Hu and Sawicki proposed a general class of broken power-law models that can satisfy the aforementioned requirements. The specific functional form of $f(R)$ suggested by Hu and Sawicki has the following form
\begin{equation}\label{husawikieqfr}
f(R)= -m^{2}\frac{c_{1}(R/m^{2})^{n}}{c_{2}(R/m^{2})^{n}+1},
\end{equation}
where $m^{2}=\kappa\bar{\rho}_{m0}/3$ defines the characteristic mass scale, and $\bar{\rho}_{m0}=\bar{\rho}(\ln a = 0)$ represents the present-day background density of matter. The parameters $c_{1}$, $c_{2}$, and $n>0$ are dimensionless free parameters that require specific determination to reproduce the expansion history and satisfy solar-system test via the chameleon mechanism \citep{2004PhRvD..69d4026K, 2007JCAP...02..022N, 2007PhRvD..76f3505F}. 

It is important to select the sign of $f(R)$ in a manner that ensures the second derivative of $f(R)$ adheres to the subsequent condition:
\begin{equation}
f_{RR}=\frac{d^{2}f(R)}{dR^{2}}>0.
\end{equation}
This requirement guarantees the stability of the solution in high-density regions when $R$ greatly exceeds $m^{2}$. Additionally, the presence of a nonzero and positive second derivative of the functional expression $f(R)$ ensures the alignment of cosmological tests derived from the model with those originating from GR In regions characterized by significantly higher curvature compared to $m^{2}$, the functional form of Hu-Sawicki model can be expanded as follows:
\begin{equation}\label{husawlim}
\lim_{m^{2}/R\rightarrow 0} f(R) \approx -\frac{c_{1}}{c_{2}}m^{2}+\frac{c_{1}}{c_{2}^{2}}m^{2}\left(\frac{m^{2}}{R}\right)^{n}
\end{equation}
Although the Hu-Sawicki model does not include a true cosmological constant, at constant $c_{1}/c_{2}$, the limiting case of $c_{1}/c_{2}^{2}\rightarrow 0$ behaves as a cosmological constant at both large-scale and local expriments. Additionally, the finite $c_{1}/c_{2}^{2}$ leads to the constant value of curvature, which remains unchanged as the matter density changes. As a result, by this choice one can have a class of models that are able to accelerate the expansion of the Universe similar to the behavior of the standard model of cosmology. Hence, one can rewrite Eq.\,(\ref{husawlim}) as follows:
\begin{equation}
f(R)=-\frac{c_{1}}{c_{2}}m^{2}-\frac{f_{R0}}{n}\frac{\bar{R}_{0}^{n+1}}{R^{n}},
\end{equation}
where $\bar{R}_{0}$ is the present-day background curvature, and $f_{R0} \equiv f_{R}(\bar{R}_{0})$. Additionally, by demanding similarity to standard model of cosmology as $|f_{R0}|\rightarrow 0$, one can deduce
\begin{equation}
\frac{c_{1}}{c_{2}}m^{2}=2\kappa\bar{\rho}_{\Lambda},
\end{equation}
where $\bar{\rho}_{\Lambda}$ can be interpreted as the background energy density of dark energy.

Eventually, from Eq.\,(\ref{potentialphi}), one can obtain the following relations
\begin{equation}
V(\phi)=\frac{Rf_{R}(1+1/n)+2\kappa\bar{\rho}_{\Lambda}}{2\kappa(1+f_{R})^{2}},
\end{equation}
\begin{equation}
V'(\phi)=\frac{R(1-\frac{n+2}{n}f_{R})-4\kappa\bar{\rho}_{\Lambda}}{\sqrt{6\kappa}(1+f_{R})^{2}},
\end{equation}
where $R/\bar{R}_{0}=(f_{R0}/f_{R})^{1/(1+n)}$. The field strength $f_{R0}$ controls the strength of the modification and is constrained by cosmological and solar-system tests \citep{2020PhRvD.102j4060D}. Different values of $f_{R0}$ have been considered in the literature, ranging from $10^{-4}$ to $10^{-8}$, e.g., \citep{2019JCAP...09..066M}. In this work, we mainly focus on the Hu-Sawicki model with $n=1$ and $|f_{R0}|=10^{-4}, 10^{-5}$, and $10^{-6}$ (labeled as $f4$, $f5$, and $f6$, respectively).
\section{Dark Matter Halo Models}\label{sec:iii}
\subsection{Halo Density Profile}
In the conventional understanding of cosmology, dark matter halos are regarded as non-linear structures that have been dispersed throughout the Universe due to the development and evolution of hierarchical structures. Initially, density fluctuations in the early Universe might have surpassed a critical threshold, leading to their collapse under the influence of self-gravitational forces, thereby making them capable of forming dark matter halos, see, e.g., \citep{2010ApJ...719..229G, 2014ApJ...788...27I, 2023PDU....4101259D}. Physically speaking, these circumstances can be described by a dimensionless parameter called density contrast, which is derived from the excursion sets theory. The density contrast is defined as $\delta (r) \equiv [\rho(r) - \bar{\rho}]/\bar{\rho}$, where $\rho(r)$ is the density of the overdense region at any given point $r$, and $\bar{\rho}$ is the average density of the background. 

In addition, the cosmological and structure formation models depend on the characteristics of the inner regions of dark matter halos. These halos' masses are governed by a radius-dependent function known as the density profile. To establish a reliable criterion for predicting the distribution of dark matter within galactic halos, various techniques such as spectroscopic observations of gravitational lensing, x-ray temperature maps, and stellar dynamics in galaxies have been utilized \citep{2005MNRAS.357...82R}. Over the past decades, both analytical approaches and numerical simulations have been employed to derive an appropriate density profile that aligns with the observed data \citep{1965TrAlm...5...87E, 1983MNRAS.202..995J, 1985MNRAS.216..273D, 1993MNRAS.265..250D, 1996ApJ...462..563N}. One of the density profiles proposed based on $N$-body simulations within the framework of cold dark matter models is known as the Navarro, Frenk, and White (NFW) density profile \citep{1996ApJ...462..563N}
\begin{equation}\label{1a}
\rho(r)=\frac{\rho_{\rm s}}{\frac{r}{r_{s}}\left(1+\frac{r}{r_{s}}\right)^{2}},
\end{equation}
where $\rho_{\rm s}$ and $r_{\rm s}$ are the scaled density and radius that vary from halo to halo.

However, using analytical methods, Einasto discovered an alternative and appropriate definition for the density profile \citep{1965TrAlm...5...87E}, which is as follows:
\begin{equation}\label{2}
\rho(r)=\rho_{\rm s} \exp \bigg\{-\frac{2}{\alpha}\left[\left(\frac{r}{r_{s}}\right)^{\alpha}-1\right]\bigg\},
\end{equation}
whre $\alpha$ is the shape parameter for the Einasto density profile. 

It is important to highlight that for both of the aforementioned expressions, one possesses
\begin{equation}
\frac{d \ln \rho(r)}{d \ln r} = -2 \hspace*{0.5cm} {\rm for} \hspace*{0.2cm} r/r_{\rm s} = 1,
\end{equation}
which implies that the logarithmic slope of the density distribution must be $-2$ at the scaled radius.

In fact, the scaled density can be specified as $\rho_{\rm s}=\rho_{\rm crit}\delta_{\rm c}$, where $\rho_{\rm crit}$ represents the critical density of the Universe at a specific redshift $z$, while $\delta_{\rm c}$ denotes the linear threshold for overdensities, which relies on the concentration parameter $C$ according to the following relation
\begin{equation}
\delta_{\rm c}=\frac{200}{3}\frac{C^{3}}{\ln(1+C)-C/(1+C)}.
\end{equation}
The concentration parameter essentially characterizes the central density of galactic halos, which is defined as the ratio between the virial radius of the halo, $r_{\rm vir}$, and its scale radius, $r_{\rm s}$. The halo virial radius covers a volume within which the average halo density is $200$ to $500$ times the critical density of the Universe. According to numerical simulations and analytical investigations, to have accurate predictions, the concentration parameter needs to dynamically evolve with mass and redshift \citep{2012MNRAS.423.3018P, 2016MNRAS.460.1214L, 2016MNRAS.456.3068O, 2017arXiv171105277O}. This aligns with the dynamics associated with the merging history of dark matter halos and their developmental pathways. This alignment stems from the fact that smaller halos have already virialized, resulting in a higher degree of concentration compared to the larger ones. Nevertheless, it has been demonstrated that within the context of $f(R)$ gravity, the concentration parameter is influenced not only by mass and redshift but also by the scalar parameter $f_{R0}$ \citep{2019MNRAS.487.1410M}. In this study, we utilize the concentration parameter formulated in \cite{2016MNRAS.456.3068O} for our computations within the framework of GR, while adopting the concentration parameter obtained from \cite{2019MNRAS.487.1410M} for our calculations involving $f(R)$ gravity. In the next section, we will delve into the statistical characteristics of dark matter halos concerning the halo mass function.
\subsection{Halo Mass Function}
The presence of dark matter halos provides a practical and fundamental way to examine the non-linear gravitational collapse in the Universe. Hence, gaining a proper statistical perspective on the mass distribution of these halos can enhance our understanding of the physics governing them. Consequently, the halo mass function has been proposed as a way to describe the mass distribution of these structures within a given volume. In simple terms, the halo mass function explains the masses of these structures that have densities higher than a certain threshold, unaffected by the Universe's expansion, and destined to collapse. As the Universe expands, the density contrast can grow to a critical point, surpassing linear regimes and entering nonlinear regimes. At this stage, overdensities detach from the expansion of the Universe, enter the turnaround phase, and undergo collapse, leading to the formation of structures.

In the standard model of cosmology based on GR, the halo mass function can be derived analytically using the framework of excursion set theory, which models the density field as a stochastic process across scales. The fundamental premise of excursion set theory is the spherical collapse model, which determines the threshold overdensity required for collapse of a spherical perturbation \citep{1974ApJ...187..425P}. In the case of an Einstein-de Sitter Universe and a spherical-collapse halo model, the threshold overdensity can be calculated as follows \citep{1997PThPh..97...49N}
\begin{equation}
\delta_{\rm sc}=\frac{3(12\pi)^{2/3}}{20}\left(1-0.0123\log\left[1+\frac{\Omega_{\rm m}^{-1}-1}{(1+z)^{3}}\right]\right),
\end{equation}
where $\Omega_{\rm m}$ represents the density parameter of matter content. Consequently, one can approximate the threshold overdensity as $1.686$ within a narrow redshift range. A key feature of the present analysis is that $\delta_{\rm sc}$ is approximately independent of halo mass $M$ due to Birkhoff's theorem, which states that the evolution of a spherical density profile is oblivious to external influences. 

In \cite{2001MNRAS.321..372J}, a suitable definition of the differential mass function has been introduced to specify different fits for dark matter halos
\begin{equation} \label{mf}
\frac{dn}{dM}=g(\sigma)\frac{\rho_{\rm m}}{M}\left| \frac{d\ln\sigma^{-1}}{d\ln M} \right|.
\end{equation}
In this context, $n(M, z)$ represents the number density of halos with a mass $M$ at redshift $z$, $\rho_m$ denotes the cosmological matter density, and $g(\sigma)$ is multiplicity function that relies on the geometry of overdensities at the time of collapse. The function $\sigma(M, z)$ signifies the linear root mean square fluctuation of overdensities on mass $M$ and at redshift $z$, precisely defined as
\begin{equation}
\sigma^{2}(M,z) \equiv \frac{1}{2\pi^{2}}\int_{0}^{\infty}P(k,z)W^{2}(k,M)k^{2}dk,
\end{equation}
where $W(k, M)$ represents the Fourier spectrum of the top-hat filter, depending on the mass $M$ and wavenumber $k$. Additionally, $P(k, z)$ stands for the redshift-dependent power spectrum of the density fluctuations.

Numerous investigations have been carried out to estimate the halo mass function through analytical methods and numerical simulations. The objective of these studies is to find the most accurate representation for cosmic observations. By incorporating a homogeneous and isotropic collapse in the standard excursion sets theory, a straightforward analytical expression for the multiplicity function emerges as follows \citep{1974ApJ...187..425P}
\begin{equation}\label{mf1}
g_{\rm ps}(\sigma) = \sqrt{\frac{2}{\pi}}\frac{\delta_{\rm
sc}}{\sigma}\exp\left(\frac{-\delta_{\rm
sc}^{2}}{2\sigma^{2}}\right),
\end{equation}
which is called the Press-Schechter (PS) mass function. The above calculation pertains to the collapse of spherically-symmetric overdensities, but the real Universe involves a much more intricate scenario. The collapse dynamics are not spherical; instead, they are triaxial, and small overdense regions need additional matter to collapse due to significant influences from the surrounding shear field \citep{2001MNRAS.323....1S}. To address this complexity, the excursion set approach adopts ellipsoidal collapse, which introduces a stochastic barrier, prompting the investigation of a general barrier.

In the case of the standard model of cosmology, a straightforward Gaussian distribution for the barrier $B$, with a mean value $\bar{B}$ that linearly changes with the variance $S\equiv \sigma^{2}(M, z)$, proves to be highly accurate in replicating the $N$-body halo mass function \citep{2011PhRvL.106x1302C, 2011PhRvD..84b3009C, 2012PhRvD..86h3011A}. Moreover, this barrier aligns with the measured overdensity required for collapse in the initial conditions \citep{2013PhRvL.111w1303A}, and it offers the advantage of having an exact solution for the Markovian multiplicity function. Consequently, the precise solution for a fixed and linear barrier can be expressed as follows:
\begin{equation}
g(\sigma) = \sqrt{\frac{2a}{\pi}} \exp\left(-\frac{a\bar{B}^2}{2S}\right) \frac{1}{\sigma} \left( \bar{B} - S \frac{d\bar{B}}{dS} \right),
\end{equation}
where $a$ is defined as $a = 1/(1 + D_{\rm B})$, and $D_{\rm B}$ is a parameter that characterizes the diffusive nature of the barrier $B(S)$ \citep{2013PhRvD..88h4015K}. 

As evident from Eq.\,(\ref{mf1}), at a fixed redshift, the mass function relies solely on the halo mass through $\sigma(M)$, and significant changes are not anticipated to occur. This mass function represents a simple model proposed for the formation of dark matter halos, known as the spherical collapse model, which often aligns with observational data. However, it exhibits quantitative deviations from numerical results at certain mass limits \citep{2001MNRAS.321..372J}.

To address this issue, one can approximate the precise solution for a generic barrier by expanding it in higher-order derivatives of the barrier. One notably successful improvement was presented by Sheth and Tormen, which is based on a more realistic model and provides a better fit to simulation results \citep{1999MNRAS.308..119S, 2001MNRAS.323....1S}. Their approach adopts an ellipsoidal collapse model with dynamical threshold density fluctuations, in contrast to the nearly global threshold used in the PS model. Actually, Sheth and Tormen introduced the idea of considering a dynamically varying threshold overdensity for ellipsoidal collapses, denoted as $\delta_{\rm ec}$, which provides a more realistic depiction of the halo mass function. By assuming zero prolateness, they calculated this quantity as
\begin{equation}
\delta_{\rm ec}(\nu)\approx\delta_{\rm sc}(1+\gamma\,\nu^{-2\beta}),
\end{equation}
where $\gamma = 0.47$, $\beta = 0.615$, and defining $\nu \equiv \delta_{\rm sc}/\sigma(M)$. It becomes evident that this quantity depends not only implicitly on the redshift but also on the mass of the structure, and it is referred to as the moving barrier. Under this assumption, one can obtain the halo mass function for ellipsoidal collapse, also known as the Sheth-Tormen (ST) mass function, which is given by
\begin{equation}\label{mf2}
g_{\rm st}(\sigma) =a \sqrt{\frac{2b}{\pi}}\frac{\delta_{\rm sc}}{\sigma}\exp\left(\frac{-a\delta_{\rm sc}^{2}}{2\sigma^{2}}\right)
\left[1+\left(\frac{\sigma^{2}}{2\delta_{\rm sc}^{2}}\right)^{p}\right],
\end{equation}
where $a=0.3222$, $b=0.707$, and $p=0.3$. It is anticipated that this mass function will exhibit greater sensitivity to changes in redshift compared to the PS mass function. 

However, in modified gravity theories, the story becomes far more complex. For example, in $f(R)$ gravity models, which introduce an additional scalar degree of freedom, Birkhoff's theorem is violated. This makes the collapse process dependent on the environment, which is known as the chameleon screening mechanism, wherein dense regions suppress the enhanced gravitational forces to some extent. Consequently, the threshold of overdensities becomes dependent on halo mass, redshift, and specific shape of the model parameter $f_{\rm R0}$, which alters the scalar field gradients \citep{2013PhRvD..88h4015K}

\begin{eqnarray} 
\delta_{\rm c}(z, M, f_{R0}) = \delta_{\rm sc}(z) \bigg[1 + \frac{b_{2}\left(m_{\rm b} - \sqrt{m_{\rm b}^2 + 1}\right)}{(1+z)^{a_{3}}} \nonumber\\
+ b_{3}(\tanh{m_{\rm b}} - 1)\bigg],\hspace*{0.5cm}
\end{eqnarray}
where
\begin{eqnarray}
m_{\rm b}(z,M,f_{R0}) = (1+z)^{a_{3}}\bigg[\log \left(\frac{M}{M_{\odot} h^{-1}}\right) \hspace*{0.5cm} \nonumber\\ 
 - \frac{m_{1}}{(1+z)^{a_{4}}}\bigg], 
\end{eqnarray}
\begin{eqnarray}
m_{1}(f_{R0}) = 1.99\log f_{R0} + 26.21, 
\end{eqnarray}
\begin{eqnarray}
b_{\rm 2} = 0.0166,
\end{eqnarray}
\begin{eqnarray}
b_3(f_{R0}) = 0.0027(2.41 - \log f_{R0}),
\end{eqnarray}
\begin{eqnarray}
a_3(f_{R0}) = 1 + 0.99\exp\left[-2.08(\log f_{R0} + 5.57)^2\right] ,
\end{eqnarray}
\begin{eqnarray}
a_4(f_{R0}) = 0.11 \{\tanh\left[0.69(\log f_{R0} + 6.65)\right] + 1\}. 
\end{eqnarray}

To account for these intricacies, the halo mass function in $f(R)$ gravity must be derived through numerical solutions of the fully nonlinear modified Einstein equations, using initial conditions based on peaks theory rather than simplistic analytical approximations. The resulting $\delta_{\rm c}(z, M, f_{R0})$ can then be incorporated into the excursion set framework, along with a drifting diffusive barrier to model the stochasticity of realistic triaxial collapse. In this way, the complex environmental couplings between the scalar field gradient and local density configurations are captured. This enables accurate predictions of the halo mass function in $f(R)$ gravity, which can exhibit significant deviations from the GR case and thus provide a signature to constrain these modified gravity models. Under there assumptions, the multiplicity function for $f(R)$ gravity is given by \citep{2013PhRvD..88h4015K}
\begin{equation} \label{mf3}
g^{f(R), \rm sx}(\sigma) \approx g^{\rm GR, sx}(\sigma) \frac{g^{f(R), \rm sk}(\sigma)}{g^{\rm GR, sk}(\sigma)},
\end{equation}
where $g^{f(R),\rm sk}(\sigma)$ and $g^{\rm GR,sk}(\sigma)$ represent multiplicity functions applying the sharp-k filter for $f(R)$ gravity and GR, respectively, and their definitions are provided as follows:
\begin{equation}
g^{\rm GR,sk}(\sigma) = \sqrt{\frac{2a}{\pi}}\frac{\delta_{\rm sc}}{\sigma} \exp\left(-\frac{a\bar{B}_{\rm GR}^2}{2S}\right),
\end{equation}
and
\begin{eqnarray}
g^{f(R),\rm sk}(\sigma) = \sqrt{\frac{2a}{\pi}}\frac{1}{\sigma} \exp\left(-\frac{a\bar{B}_{f(R)}^2}{2S}\right) \hspace*{0.8cm }\nonumber\\
\times \left( \bar{B}_{f(R)} - S \frac{d\bar{B}_{f(R)}}{dS} \right).
\end{eqnarray}
The value of $a\simeq 0.714$ has been considered in the method developed in \cite{2013PhRvD..88h4015K}. In above relations, collapse barriers have the following form
\begin{equation}
\bar{B}_{\rm GR}(\sigma) = \delta_{\rm sc} + \beta S,
\end{equation}
\begin{equation} 
\bar{B}_{f(R)}(\sigma) = \delta_{\rm c}(z, M, f_{R0}),
\end{equation}
where $\beta$ characterizes the linear drift of the barrier height as a function of the variance S. The value of $\beta$ needs to be determined by calibrating to $N$-body simulations or models of ellipsoidal collapse. Typical values are $\beta \simeq 0.1-0.2$ for a standard model of cosmology. In the context of $f(R)$ gravity, the drift parameter $\beta$ would likely be slightly modified from its GR value, since the dynamics of aspherical collapse can be altered. However, as a first approximation, one can assumes the GR value of $\beta$ in deriving the $f(R)$ halo mass function.

Moreover, $g^{\rm GR, sx}(\sigma)$ refers to the multiplicity function calculated in the context of GR using a sharp-x filtering method, which establishes a connection between halo mass and variance. Its complete expression is \citep{2013PhRvD..88h4015K}
\begin{eqnarray} 
g^{\rm GR, sx}(\sigma)= g_0(\sigma) + g_{1,\beta=0}(\sigma) + g_{1,\beta(1)}(\sigma)+ g_{1,\beta(2)}(\sigma),\nonumber\\
\end{eqnarray}
where
\begin{eqnarray}
g_0(\sigma) =\frac{\delta_{\rm sc}}{\sigma} \sqrt{\frac{2a}{\pi}} \exp\left(-\frac{a\delta_{\rm sc}^2}{2S}\right),
\end{eqnarray}
\begin{eqnarray}
g_{1,\beta= 0}(\sigma) =- \frac{\tilde{\kappa}\delta_{\rm sc}}{\sigma} \sqrt{\frac{2a}{\pi}} \bigg[ \exp\left(-\frac{a\delta_{\rm sc}^2}{2S}\right)\hspace*{0.8cm} \nonumber\\
- \frac{1}{2}\Gamma\left(0, \frac{a\delta_{\rm sc}^2}{2S}\right) \bigg],
\end{eqnarray}
\begin{eqnarray}
g_{1,\beta^{(1)}}(\sigma) = -a\delta_{\rm sc}\beta \left[ \tilde{\kappa} \text{Erfc}\left(\frac{\delta_{\rm sc}a^{2}}{\sqrt{2S}}\right) + g_{1,\beta=0}(\sigma) \right],
\end{eqnarray}
\begin{eqnarray}
g_{1,\beta^{(2)}}(\sigma) = -a\beta\left[\frac{1}{2}\beta S g_{1,\beta=0}(\sigma) + \delta_{\rm sc} g_{1,\beta^{(1)}}(\sigma)\right].
\end{eqnarray}
In the above relations, $\tilde{\kappa} \simeq 0.65\,S/\delta_{\rm sc}^2$, $\Gamma(x, y)$ is the incomplete gamma function, and $\text{Erfc}(x)$ is the complementary error function. By following all the outlined procedures, the halo mass function can be derived within the framework of the $f(R)$ gravity. Clearly, this mass function relies not only on mass and redshift but also on the field strength $f_{R0}$. Consequently, variations in $f_{R0}$ are anticipated to lead to changes in the halo mass function in $f(R)$ gravity.

\begin{figure}[t!] 
\begin{minipage}{1\linewidth}
\includegraphics[width=1\textwidth]{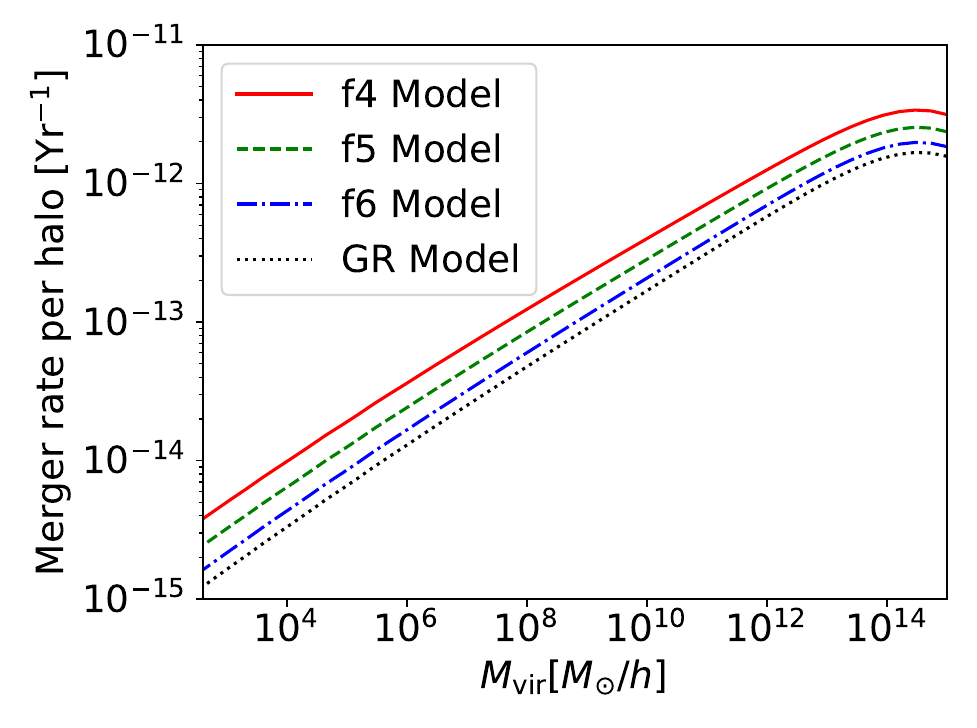}
\caption{The merger rate of PBHs within each halo as a function of halo mass for Einasto density profile. Solid, dashed, and dot-dashed lines exhibit this relation for dark matter halos in $f(R)$ gravity, with $|f_{R0}| = 10^{-4}$, $10^{-5}$, and $10^{-6}$, respectively. While the dotted line indicates it for dark matter halos in GR.}
\label{fig:per_year}
\end{minipage}
\end{figure}
\section{Merger rate of PBHs} \label{sec:iv}
As previously mentioned, PBHs are distributed randomly throughout the Universe, which means they can encounter each other and form binary systems. In this section, our objective is to compute the merger rate of PBHs within the context of $f(R)$ gravity and compare the results with those obtained from GR. To achieve this, we have previously discussed the essential tools for modeling dark matter halos.

Let us consider two PBHs with masses $m_1$ and $m_2$ and a relative velocity at large separation $v_{\rm rel} = |v_1-v_2|$ that coincidentally encounter each other within a dark matter halo. Under these conditions and according to Keplerian mechanics, the maximum gravitational radiation is emitted at the periastron $r_{\rm p}$. If the emitted gravitational energy is greater than the kinetic energy of the system, PBHs will become gravitationally bound and form binary systems. As a consequence of this situation, there is an upper limit to the periastron distance \citep{123456789p}
\begin{equation}\label{priastronmp}
r_{\rm mp} = \left[\frac{85\pi \sqrt{2} G^{7/2}m_{1}m_{2}(m_{1}+m_{2})^{3/2}}{12c^{5}v_{\rm rel}^{2}}\right]^{2/7},
\end{equation}
where $G$ represents the gravitational constant and $c$ stands for the velocity of light. Moreover, within the Newtonian limit, a connection between the impact parameter and the periastron is indicated by the following relation \citep{2018CQGra..35f3001S}
\begin{equation}
b^{2}(r_{\rm p})=\frac{2G(m_{1}+m_{2})r_{\rm p}}{v_{\rm rel}^{2}}+r_{\rm p}^{2}.
\end{equation}

When considering scenarios where the tidal forces of surrounding black holes rarely result in head-on collisions, it can be expected that the gravitational interaction between the two PBHs leads to the formation of a binary system with maximum eccentricity. On the other hand, under the strong gravitational focusing limits, it can be roughly implied that the tidal forces of the surrounding black holes cannot disturb the orbital parameters of the formed binaries.

Note that PBH binaries forming within dark matter halos have merger times ranging from hours to kiloyears, with dissipative two-body encounters resulting in much shorter merger times than the age of the Universe. On the other hand, binaries formed through nondissipative three-body encounters typically have longer merger times, making them less likely to significantly contribute to the population of BH mergers observed by LIGO-Virgo-KAGRA detectors. Nevertheless, there might be specific circumstances where the merger rate of binaries formed through three-body encounters is likely significant. In this work, we mainly focus on those binaries that can be formed through two-body encounters. Therefore, the cross-section for binary formation can be computed using the following relation \citep{1989ApJ...343..725Q, 2002ApJ...566L..17M}
\begin{equation}\label{xicrosssec}
\xi(m_{1}, m_{2}, v_{\rm rel})=\pi b^{2}(r_{\rm p, max})\simeq \frac{2\pi G(m_{1}+m_{2})r_{\rm p, max}}{v_{\rm rel}^{2}}.
\end{equation}
As a result, when Eq.\,(\ref{priastronmp}) is inserted into Eq.\,(\ref{xicrosssec}), it leads to an explicit expression for the cross-section for the binary formation
\begin{equation}
\xi \simeq 2\pi \left(\frac{85\pi}{6\sqrt{2}}\right)^{2/7}\frac{G^{2}(m_{1}+m_{2})^{10/7}(m_{1}m_{2})^{2/7}}{c^{10/7}v_{\rm rel}^{18/7}}.
\end{equation}
This relation is derived using strong limit gravitational focusing, where the tidal forces from neighboring compact objects on the binary system can be disregarded, i.e., $r_{\rm p}\ll b$.

\begin{figure}[t!] 
\begin{minipage}{1\linewidth}
\includegraphics[width=1\textwidth]{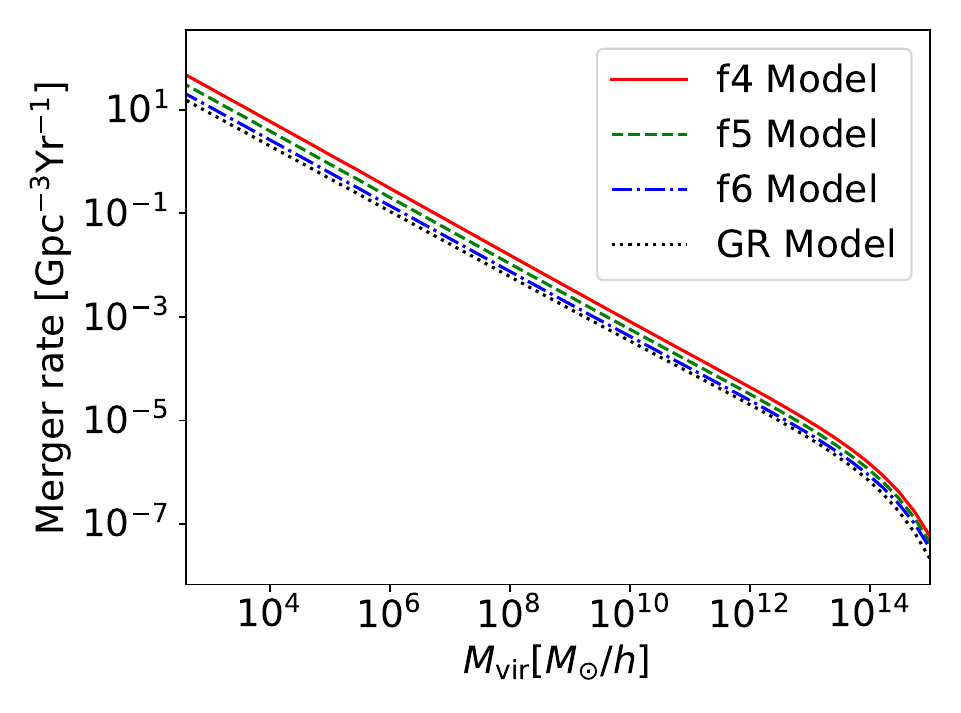}
\caption{Total merger event rate of PBHs per unit source time and per unit comoving volume as a function of halo mass for Einasto density profile. Solid, dashed, and dot-dashed lines exhibit this relation for dark matter halos in $f(R)$ gravity, with $|f_{\rm R0}| = 10^{-4}$, $10^{-5}$, and $10^{-6}$, respectively. While the dotted line indicates it for dark matter halos in GR.}
\label{fig:tot}
\end{minipage}
\end{figure}

In this study, we focus on events that align with the sensitivity of LIGO-Virgo-KAGRA detectors. Therefore, our analysis is limited to cases where $m_{1}=m_{2}=M_{\rm PBH}$. With this constraint in mind, the binary formation rate within a galactic halo can be calculated as follows \citep{2016PhRvL.116t1301B}
\begin{equation}
\Gamma(M_{\rm vir})=\int_{0}^{r_{\rm vir}}2\pi r^{2}\left(\frac{f_{\rm PBH}\rho_{\rm halo}}{M_{\rm PBH}}\right)\langle\xi v_{\rm rel}\rangle dr,
\end{equation} 
where $0 < f_{\rm PBH} \leq 1$ indicates the proportion of PBHs contributing to dark matter. Additionally, $\rho_{\rm halo}$ denotes the halo density profile, and the angle bracket represents an average over the distribution of PBH relative velocities within the galactic halo.

Furthermore, the halo virial mass, representing the mass enclosed within the virial radius, is determined using the following equation
\begin{equation}
M_{\rm vir}=\int_{0}^{r_{\rm vir}} 4\pi r^{2} \rho_{\rm halo}(r)dr,
\end{equation}
which can be computed, taking into account the NFW and Einasto density profiles.

\begin{figure}[t!] 
\begin{minipage}{1\linewidth}
\includegraphics[width=1\textwidth]{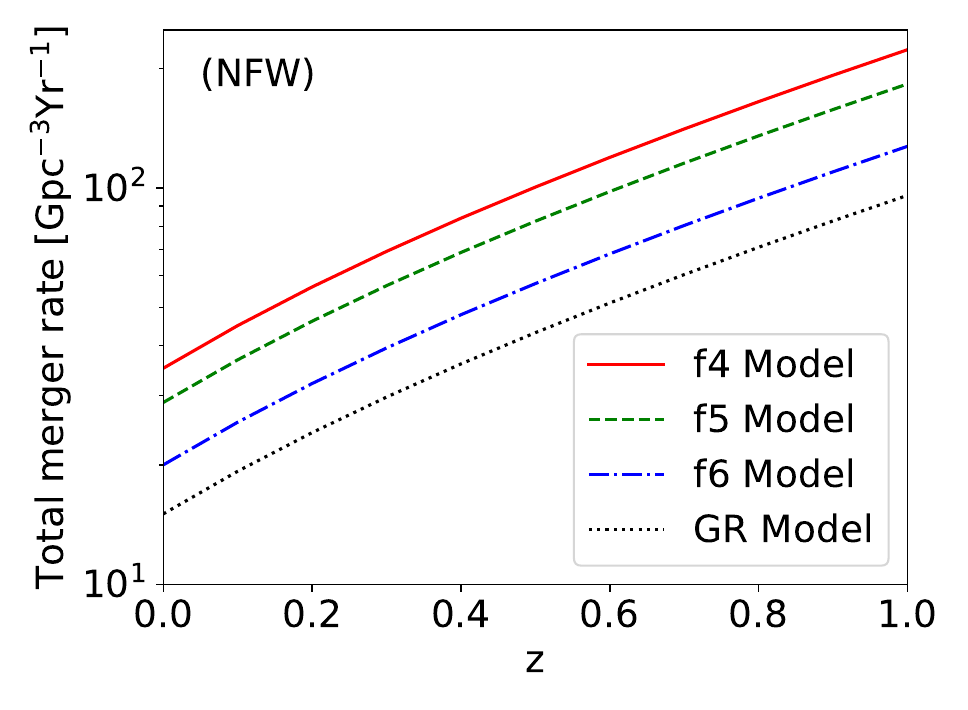}\\~\\
\includegraphics[width=1\textwidth]{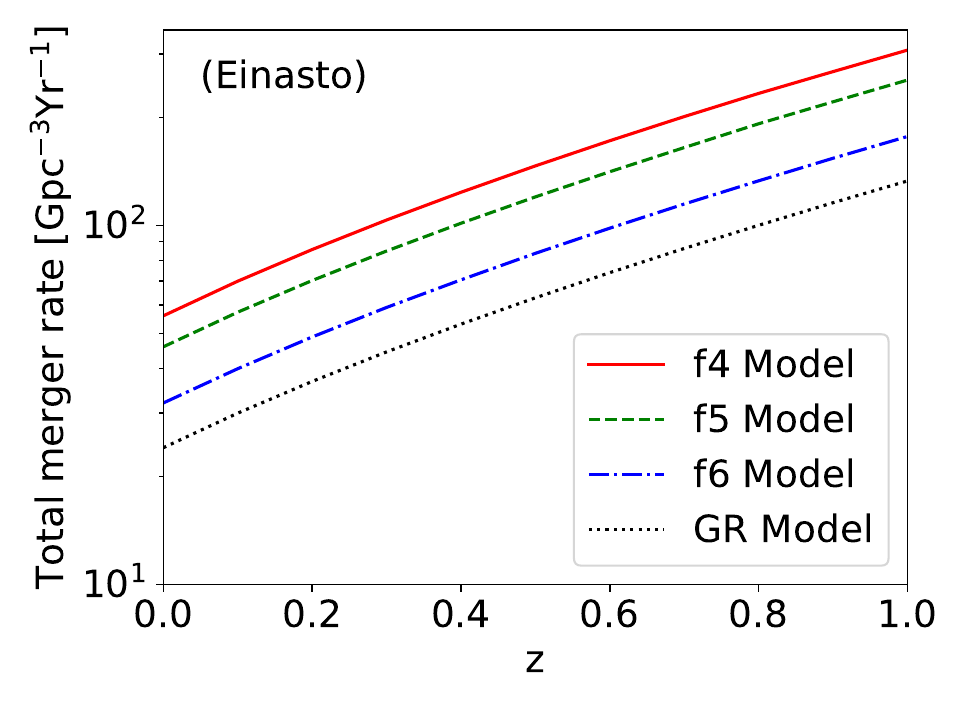}
\caption{Redshift evolution of the total merger rate of PBHs per unit source time and per unit comoving volume for (top) NFW and (bottom) Einasto density profiles. Solid, dashed, and dot-dashed lines exhibit this relation for dark matter halos in $f(R)$ gravity, with $|f_{\rm R0}| = 10^{-4}$, $10^{-5}$, and $10^{-6}$, respectively. While dotted lines indicate it for dark matter halos in GR.}
\label{fig:red}
\end{minipage}
\end{figure}

The velocity dispersion of dark matter particles in galactic halos is another essential element in computing the merger rate of PBHs. In this regard, it has been revealed that the velocity dispersion profiles in $f(R)$ gravity are similar to those in the standard model of cosmology \citep{2015PhRvL.115g1306H}. Thus, it is feasible to utilize, with reasonable accuracy, the velocity dispersion relation of dark matter particles obtained in \cite{2012MNRAS.423.3018P} for both $f(R)$ gravity and GR
\begin{equation}
v_{\rm disp}=\frac{v_{\rm max}}{\sqrt{2}}=\sqrt{\frac{GM(r<r_{\rm max})}{r_{\rm max}}}.
\end{equation}
where $v_{\rm max}$ represents the maximum velocity within a radius of $r_{\rm max}$. Additionally, we require that the Maxwell-Boltzmann distribution, truncated at the virial velocity, governs the probability distribution function of relative velocities among PBHs in the galactic halos
\begin{equation} 
P(v_{\rm PBH}, v_{\rm disp})=\mathcal{J}_{0}\left[\exp\left(-\frac{v_{\rm PBH}^{2}}{v_{\rm disp}^{2}}\right)-\exp\left(-\frac{v_{\rm vir}^{2}}{v_{\rm disp}^{2}}\right)\right],
\end{equation}
where $\mathcal{J}_{0}$ is determined by fulfilling the requirement that $4\pi \int_{0}^{v_{\rm vir}}P(v)v^{2}dv = 1$, and $v_{\rm PBH}=v_{\rm rel}$.

In the realm of PBH binary formation, two distinct mechanisms come into play, coexisting harmoniously in different epochs of the Universe \citep{2018CQGra..35f3001S}. This study delves into the formation of PBH binaries within dark matter halos during the late-time Universe. Theoretical forecasts from this process even propose the intriguing possibility that the vast majority of dark matter might consist of PBHs \citep{2016PhRvL.116t1301B, 2021PhRvD.103l3014F, 2022PhRvD.105d3525F, 2023PhRvD.107f3507F}.

However, during the early Universe, the initial clustering of PBHs could lead them to break away from the Hubble flow, giving rise to binary formations \citep{2017PhRvD..96l3523A}. These PBH binaries, emitting GWs constantly, gradually reduce in size and eventually merge. Nonetheless, the merger process is not without its challenges, as tidal forces from surrounding PBHs might disrupt some of these binaries before they come to merge \citep{2018PhRvD..98b3536K, 2019JCAP...02..018R}. 

The merger time for PBH binaries formed in the early Universe hinges on their orbital parameters, which, due to the random distribution of PBHs during that era, can vary significantly at the time of their binary formation. Consequently, a diverse array of scenarios unfolds: some PBH binaries have already merged, others are destined to merge in the current Universe, and still others await their fate in the future. It is in this context that LIGO-Virgo-KAGRA observations can be elucidated, showcasing how PBHs constitute only a tiny fraction of dark matter \citep{2020PhRvD.102l3524H, 2021JCAP...03..068H, 2022PhLB..82937040C}. It is crucial to note that the journey of PBHs in the cosmic symphony continues to enthrall scientists, paving the way for captivating discoveries yet to come. Hence, both mechanisms remain valid for describing black hole binary mergers, and the predictions concerning the abundance of PBHs are still under scrutiny by the LIGO-Virgo-KAGRA detectors. In this context, we minimally assume the contribution of PBHs to dark matter as $f_{\rm PBH} = 1$ when calculating the merger rate of PBHs. I also set the mass of PBHs to be $M_{\rm PBH}=30\,M_{\odot}$. However, we will later extend our analysis to encompass the changes in both mass and fraction of PBHs.

In Fig. \ref{fig:per_year}, we have presented the merger rate of PBHs within each halo as a function of the halo mass for three models of $f(R)$ gravity, i.e., $f4$, $f5$ and $f6$, and compared them with the merger rate obtained for GR while considering the Einasto density profile. It is clear from the figure that the merger rate of PBHs within each halo, in comparison to the outcome derived from GR, is higher for all the examined models under $f(R)$ gravity. This can be attributed to the effect of nonlinear dynamics considered for density fluctuations, $\delta_{\rm c}(z, M, f_{R0})$, and formation and evolution of halo structures in $f(R)$ gravity, which appears in all stages of calculations. An additional intriguing finding emerges as we observe a gradual reduction in the influence of the field strength-dependent dynamics in density fluctuations, moving from $f4$ to $f6$. This can be attributed to the fact that the merger rate of PBHs, as dark matter candidates, is directly shaped not only by the dynamics of density fluctuations but also by the field strength $f_{R0}$.

Furthermore, the LIGO-Virgo-KAGRA detectors are designed to record and process the accumulated merger rate of black holes. Therefore, it is necessary to calculate the overall merger rate of PBHs per unit volume and time. To achieve this, one needs to combine the halo mass function, $dn/dM_{\rm vir}$, with the merger rate per halo, $\Gamma(M_{\rm vir})$,
\begin{equation}\label{totmerrate}
\mathcal{R}=\int_{M_{\rm c}}\frac{dn}{dM_{\rm vir}} \Gamma(M_{\rm vir})dM_{\rm vir},
\end{equation}
The significance of the upper limit of integration in determining the merger rate of PBHs can be easily dismissed. This is due to the decreasing nature of the halo mass function, which causes the contribution of PBH merger rate to diminish exponentially as the halo mass increases. This behavior aligns with the hierarchical dynamics of halo formation, where low-mass halos have higher dark matter density than high-mass halos because they have already undergone virialization. As a result, the lower limit of integration holds greater importance in this analysis. Referring to arguments presented in \cite{2021PhRvD.103l3014F, 2022PhRvD.105d3525F}, when calculating the merger rate of PBHs with a mass of $30\,M_{\odot}$, the lower limit for halo mass needs to be set at $M_{\rm c} \simeq 400\,M_{\odot}$. This implies that signals from dark matter halos with masses below $400\,M_{\odot}$ are expected to be negligible. Additionally, in this analysis, we employ Eq.\,(\ref{mf2}) to represent the halo mass function for GR, while we use Eq.\,(\ref{mf3}) for the halo mass function in the context of $f(R)$ gravity.

\begin{figure}[t!] 
\begin{minipage}{1\linewidth}
\includegraphics[width=1\textwidth]{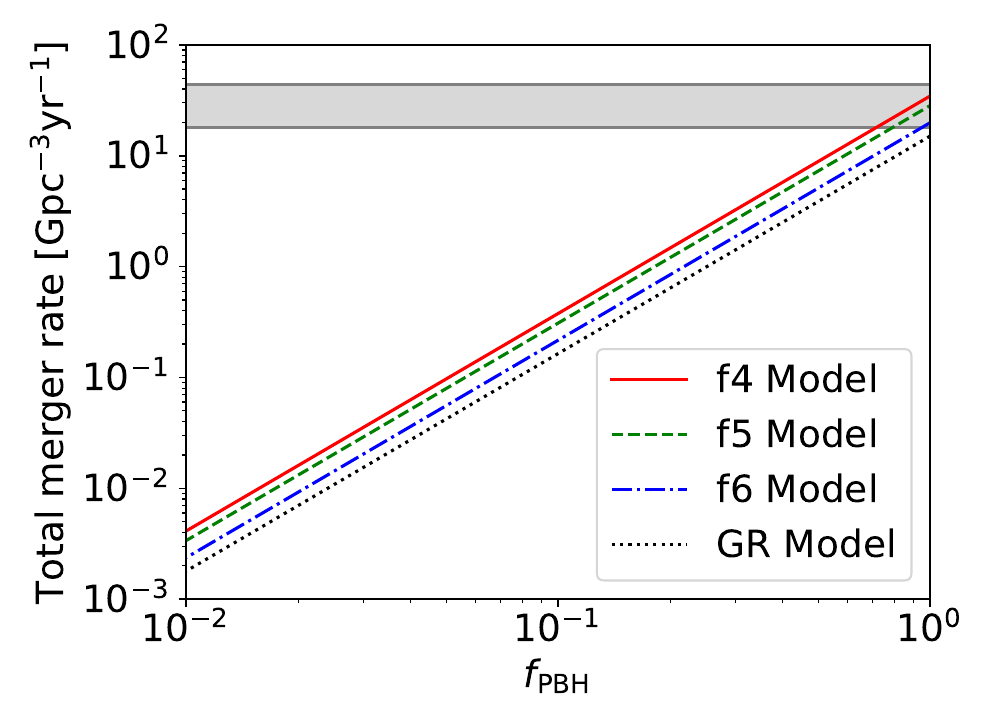}\\~\\
\includegraphics[width=1\textwidth]{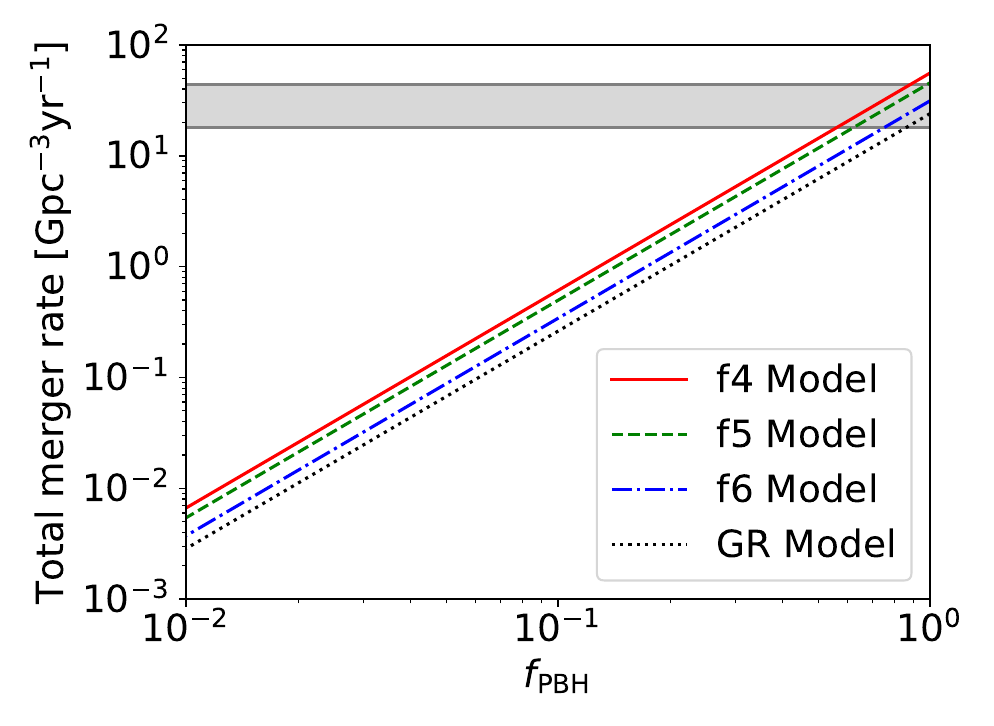}
\caption{Total merger event rate of PBHs with respect to the PBH fraction for (top) NFW and (bottom) Einasto density profiles. Solid, dashed, and dot-dashed lines exhibit this relation for dark matter halos in $f(R)$ gravity, with $|f_{\rm R0}| = 10^{-4}$, $10^{-5}$, and $10^{-6}$, respectively. While the dotted line indicates it for dark matter halos in GR. The mass of involving PBHs has been considered as $M_{\rm PBH}=30 M_{\odot}$. Also, The shaded band represents the total merger rate of black holes estimated by the LIGO-Virgo-KAGRA detectors during the second half of the third observing run (O3b), i.e., $(17.9\mbox{-}44)\,{\rm Gpc^{-3} yr^{-1}}$.}
\label{fig:frac}
\end{minipage}
\end{figure}

In Fig.\,\ref{fig:tot}, we have displayed the merger rate of PBHs per unit time and volume in three $f(R)$ gravity models, i.e., $f4$, $f5$ and $f6$, and compared them with the corresponding results obtained from GR while taking into account the Einasto density profile. These calculations are performed for the present-time Universe. As it is evident from the inset figure, smaller-mass halos continue to play a more significant role in driving the merger rate of PBHs compared to the larger halos. This is a direct consequence of the higher density of dark matter particles within subhalos, as previously discussed, creating a higher concentration in subhalos compared to the host halos. It is important to highlight that the merger rate of PBH binaries can be precisely quantified through the integration across the area below the curves. Furthermore, it can be deduced that across all the considered models within $f(R)$ gravity, the merger rate of PBHs surpasses that obtained from GR. Also, the direct dependence of the merger rate of PBHs with the field strength $f_{R0}$ is evident. Evidently, as the field strength weakens, the merger rate of PBHs for $f(R)$ gravitaty progressively converge towards the findings obtained through GR. This implies that by constraining the value of field strength and comparing the predictions from the present analysis with GW data, it is possible to potentially constrain the abundance of PBHs, which in turn introduces a novel method.

\begin{figure*}[ht!] 
\gridline{\fig{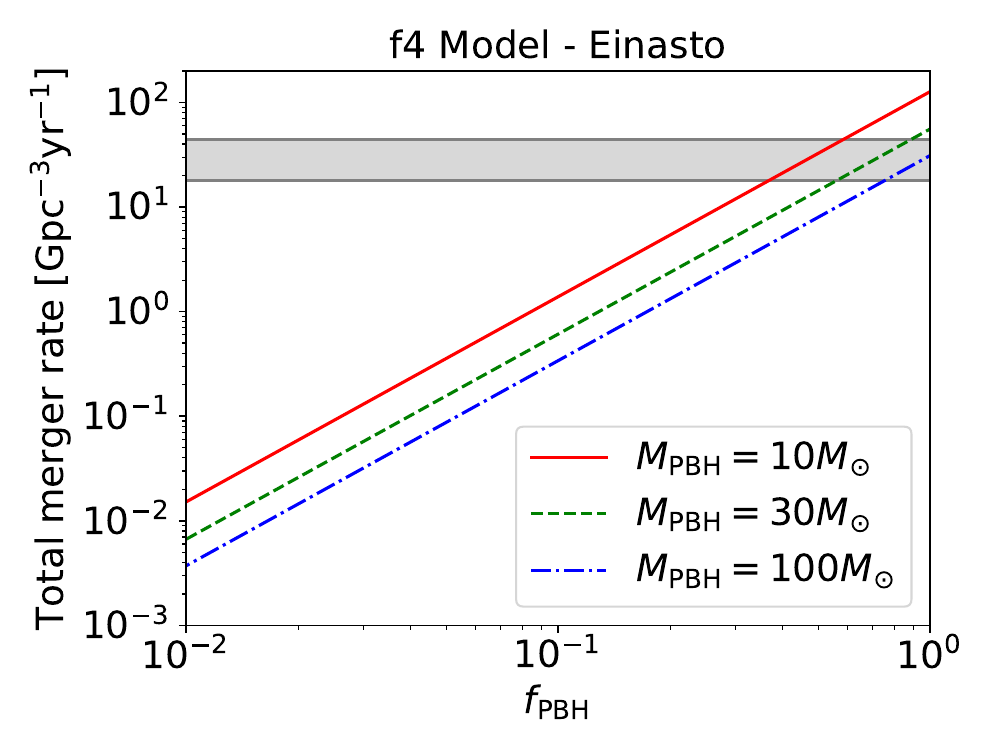}{0.5\textwidth}{}
\fig{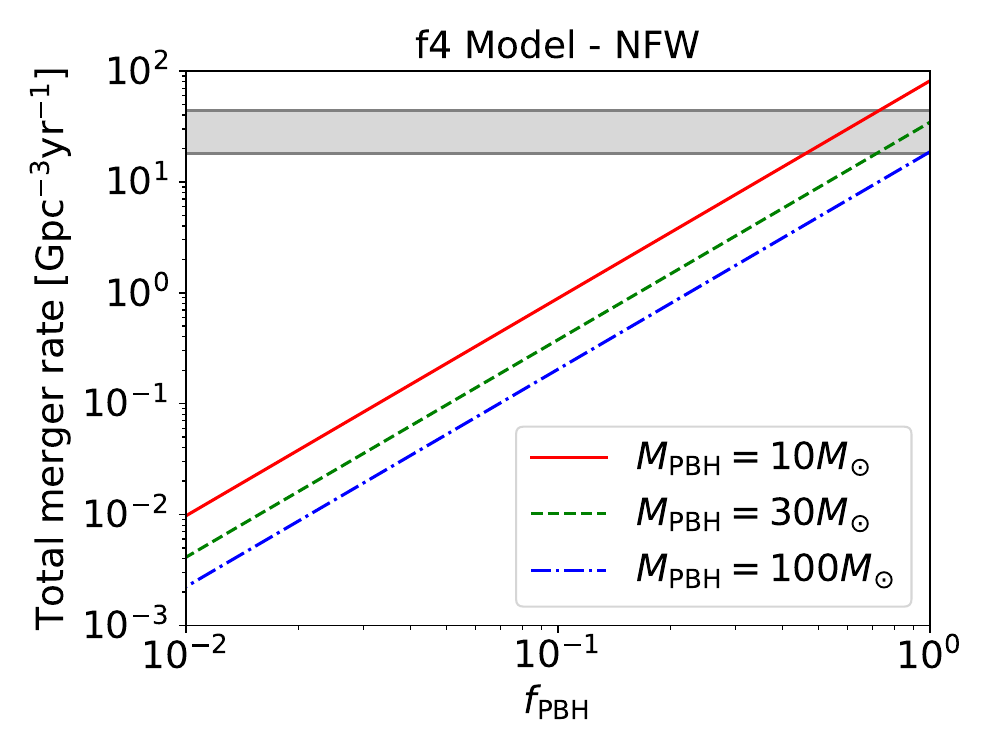}{0.5\textwidth}{}
}
\gridline{\fig{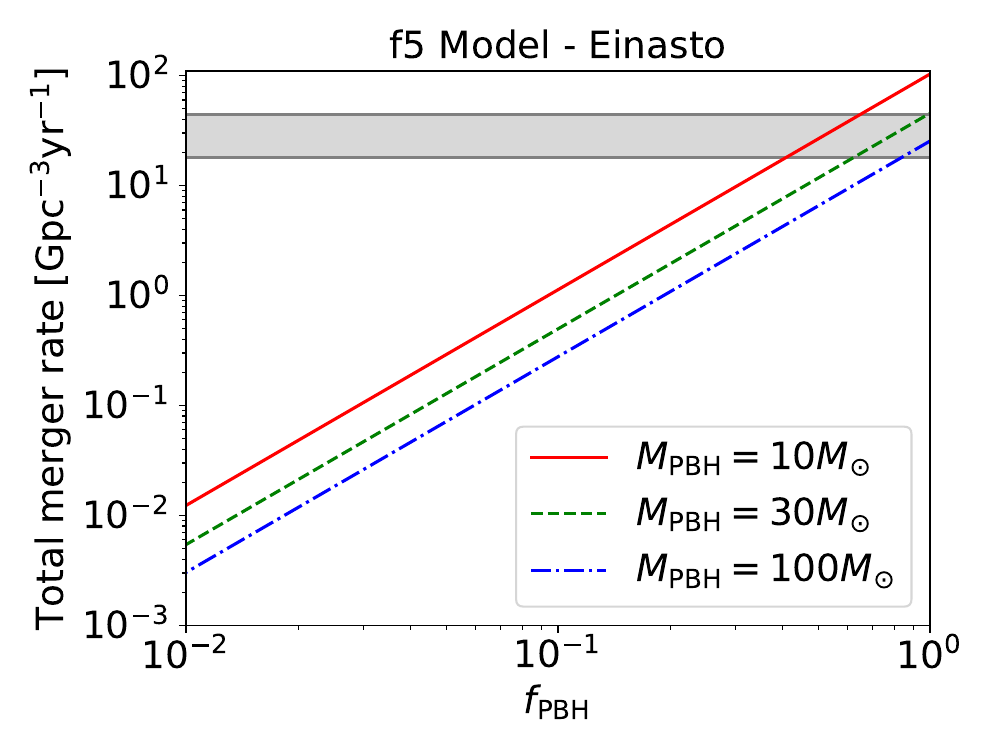}{0.5\textwidth}{}
\fig{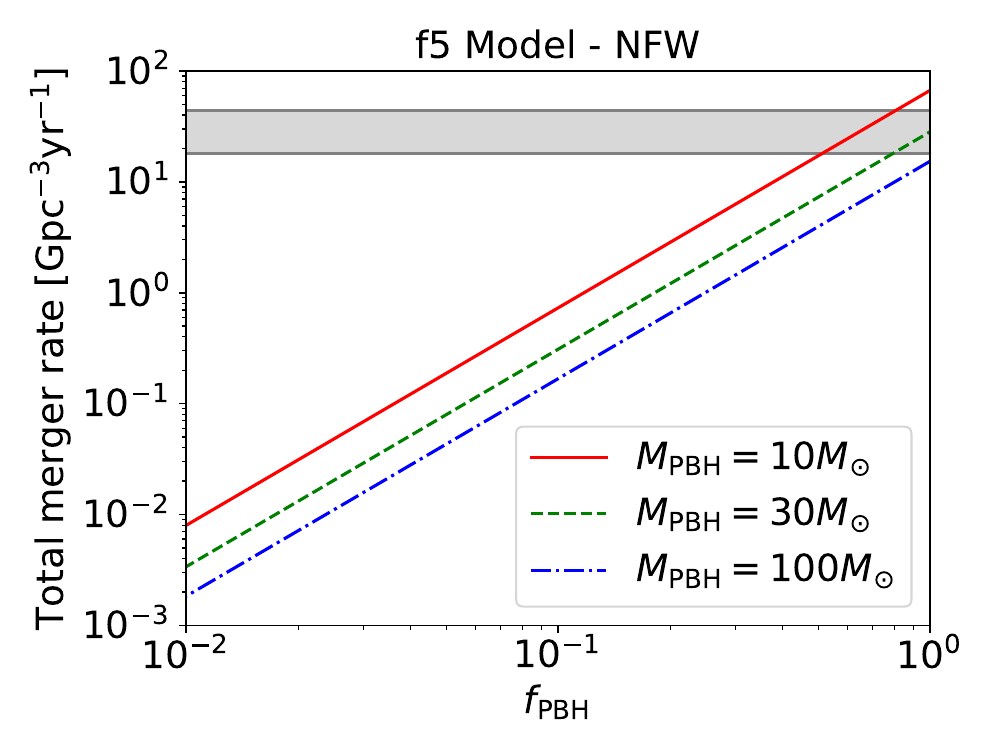}{0.5\textwidth}{}
}
\gridline{\fig{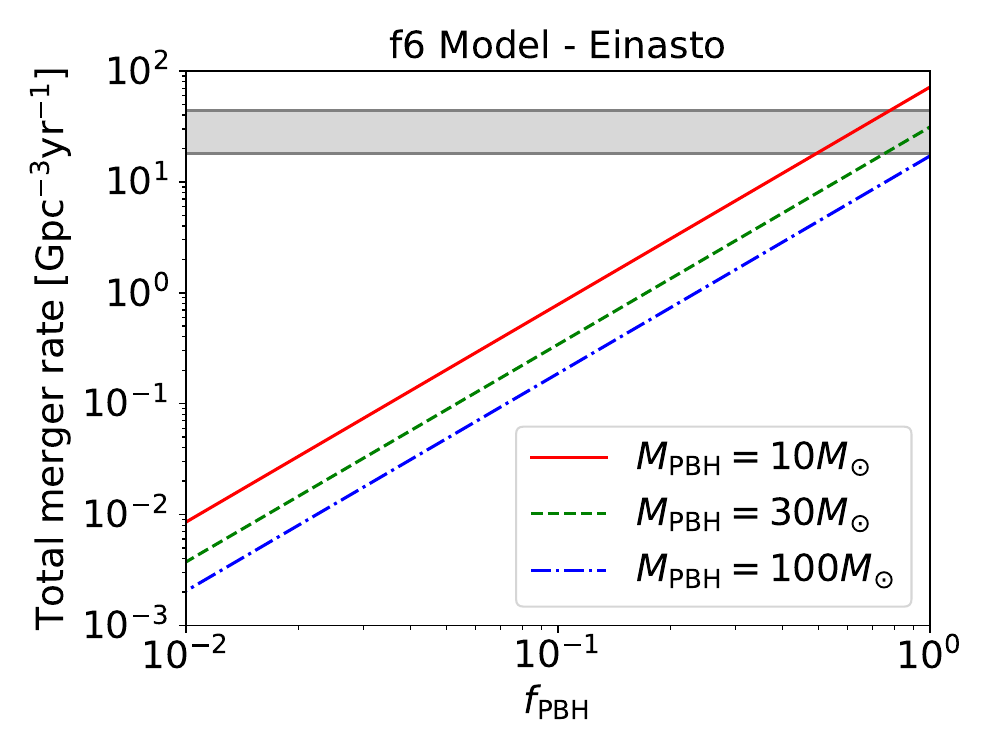}{0.5\textwidth}{}
\fig{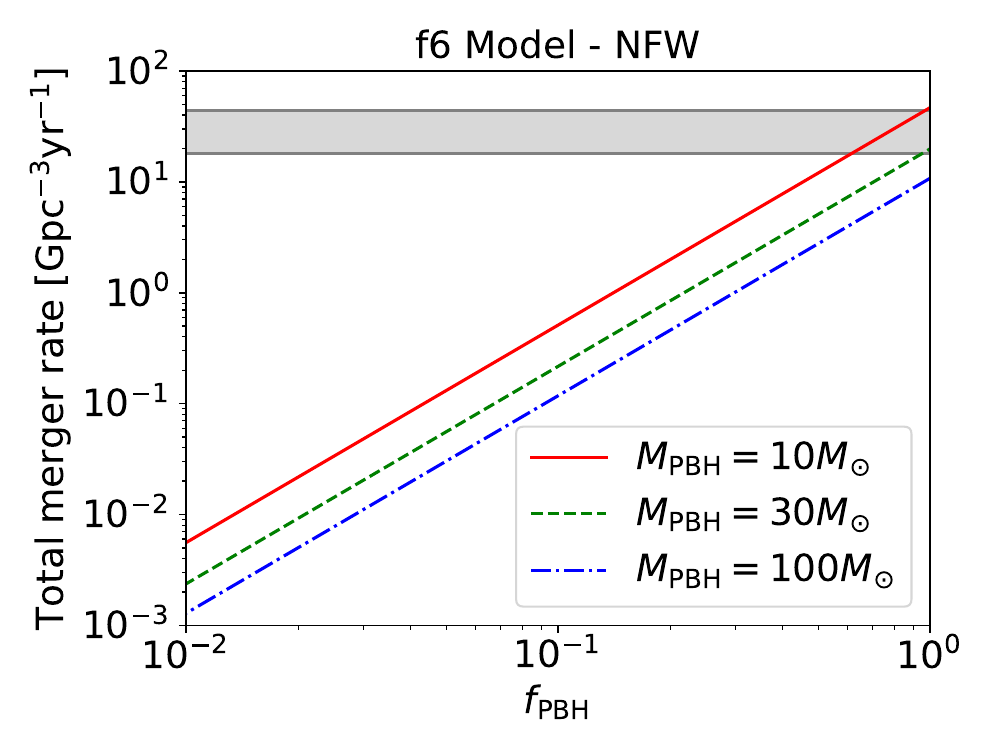}{0.5\textwidth}{}
}
\caption{Total merger event rate of PBHs with respect to their fraction for three models of $f(R)$ gravity, i.e., $f4$, $f5$, and $f6$, while taking into account the PBH masses to be $M_{\rm PBH} = 10$, $30$, and $100\,M_{\odot}$. The shaded band represents the total merger rate of black holes, as estimated by the LIGO-Virgo-KAGRA detectors during the second half of the third observing run (O3b), within the range of $(17.9$-$44)\,{\rm Gpc^{-3} yr^{-1}}$. Both NFW and Einasto density profiles have been incorporated.}
\label{fig:mass_frac}
\end{figure*}

One of the primary purposes of this analysis is to compute the merger rate of PBHs in the distant Universe, allowing for a meaningful comparison with the merger events detected by the LIGO-Virgo-KAGRA observatories. This endeavor aims to offer theoretical forecasts based on the gravitational models proposed in this research, shedding light on the forthcoming landscape of GW phenomena and the advancement of these detection instruments. Through our analysis, we aim to illustrate the redshif evolution of the merger rate of PBHs. On the other hand, with the present sensitivity of GW detectors, they can detect merger events occurring within a comoving volume of $50\,{\rm Gpc^{3}}$, equivalent to a redshift of approximately $z \sim 0.75$ \citep{2021PhRvX..11b1053A, 2021arXiv211103606T}. This raises an intriguing question of computing the redshift evolution of the merger rate of PBHs. It should be emphasized that Eq.\,(\ref{totmerrate}), which incorporates the halo mass function and the concentration parameter, is sensitive to changes in redshift. To address this objective, we have presented in Fig.\,\ref{fig:red} the redshift evolution of the merger rate of PBHs for three $f(R)$ gravity models, i.e., $f4$, $f5$ and $f6$. Additionally, we have compared these results with the corresponding outcomes for GR while considering both NFW and Einasto density profiles. The direct relation between the merger rate PBHs and redshift is evident. This can be explained by considering the impact of hierarchical dynamics and the structure of halo mergers. This suggests that during earlier periods characterized by higher redshifts, there may have been a greater abundance of subhalos. Consequently, in higher redshifts, PBHs merged at a more accelerated rate compared to the present-time Universe. The findings indicate that the redshift evolution of the merger rate of PBHs within $f(R)$ gravity models exceeds that obtained from the framework of GR. This implies that, assuming the credibility of $f(R)$ gravity, the merger rate of PBHs will be amplified over time. Furthermore, it is evident that the impact of the field strength $f_{R0}$ on enhancing the merger rate of PBHs will endure during the late-time Universe.

\begin{table*}[t!] 
\caption{Total merger rate of PBHs in the context of GR and $f(R)$ gravity with field strengths $f4$, $f5$, and $f6$ as a function of different PBH masses, i.e., $M_{\rm PBH} = 10, 20, 30, 50,$ and $100\,M_{\odot}$, while considering NFW and the Einasto density profiles, at the present-time Universe $(z=0)$.}
\centering
\begin{tabular}{c|c|c|c|c|c}
\hline
\hline
PBH Mass $(M_{\odot})$ & Density Profile &$\mathcal{R}_{\rm GR} (\rm Gpc^{-3}\rm yr^{-1})$&$\mathcal{R}_{\rm f4}(\rm Gpc^{-3}\rm yr^{-1})$ & $\mathcal{R}_{\rm f5} (\rm Gpc^{-3}\rm yr^{-1})$& $\mathcal{R}_{\rm f6} (\rm Gpc^{-3}\rm yr^{-1})$\\ [0.5ex]
\hline
10 & NFW & $33.4$ & $81.8$ & $66.9$ & $46.8$\\
10 & Einasto & $51.2$ & $127.1$ & $103.4$& $71.8$\\
\hline
20 & NFW & $22.3$ & $51.5$ & $42.2$& $29.5$\\
20 & Einasto & $35.0$ & $81.7$ & $66.6$& $46.3$\\
\hline
30 & NFW & $15.0$ & $34.6$ & $28.3$& $19.8$\\
30 & Einasto & $24.0$ & $55.8$ & $45.5$& $31.3$\\
\hline
50 & NFW & $11.4$ & $26.3$ & $21.6$& $15.1$\\
50 & Einasto & $18.6$ & $43.1$ & $35.2$& $24.5$\\
\hline
100 & NFW & $8.1$ & $18.7$ & $15.3$& $10.8$\\
100 & Einasto & $13.4$ & $31.1$ & $25.4$& $17.1$\\
\hline
\hline
\end{tabular}
\label{table1}
\end{table*}

Hereafter, our attention will be focused on the expected PBH fraction, $f_{\rm PBH}$, which comes from the $f(R)$ gravity models. The issue of the fraction of PBHs has been a significant concern from the outset of the development of the PBH scenario. Furthermore, a crucial constraint placed on PBHs pertains to their abundance within the Universe during late times. Nowadays, the abundance fractions of many mass ranges of PBHs have been strongly constrained using observational and nonobservational methods \citep{2021RPPh...84k6902C, 2023arXiv230603903C}. However, a specific mass interval of PBHs, referred to as asteroid mass PBHs, i.e., those with masses around $10^{-17} M_{\odot} \leq M_{\rm PBH} \leq 10^{-12} M_{\odot}$ \citep{2019JCAP...08..031M, 2020PhRvD.101f3005S, 2021PhRvD.104b3516R, 2022arXiv220814279G}, has yet to be definitively constrained, and could potentially account for a substantial portion of the dark matter composition. On the othe hand, comparing the PBH merger rate obtained from each theoretical model with the estimated one via the LIGO-Virgo-KAGRA detectors can potentially be one of the best references for validating that model. In Fig.\,\ref{fig:frac}, based on NFW and Einasto density profiles, we have plotted the total merger rate of PBHs as a function of $f_{\rm PBH}$ for the models of $f(R)$ gravity and compared the results with those extracted from GR. Additionally, the shaded band represents the total merger rate of black holes estimated by the LIGO-Virgo-KAGRA detectors during the latest observing run, O3b, which is $(17.9\mbox{-}44)\, {\rm Gpc^{-3}yr^{-1}}$ \citep{2021arXiv211103606T}. It is evident from the figure that the merger rate of PBHs for both gravitational models is inversely proportional to their masses but directly proportional to their fractions. This is due to the fact that the number density of PBHs changes inversely with their masses. It can also be inferred that, compared to the corresponding results derived from GR, the models of $f(R)$ gravity satisfy the constraints stemming from GW data for relatively lower values of the fraction of PBHs. In addition, the direct effect of field strength $f_{R0}$ from $f(R)$ gravity in imposing stringent constraints on the fraction of PBHs is evident in the present analysis.

Up until now, we have assumed in our analysis that the mass of the involved PBHs is $30\,M_{\odot}$ and that they can contribute maximally to dark matter. However, it is of interest to calculate the merger rate of PBHs based on different assumptions regarding their fractions and masses. To this end, in Fig.\,\ref{fig:mass_frac}, we have depicted the total merger rate of PBHs as a function of $10,M_{\odot} \leq M_{\rm PBH} \leq 100,M_{\odot}$ and $f_{\rm PBH}$ for $f(R)$ gravity. In this calculation, NFW and Einasto density profiles have been incorporated, considering three values of field strength: $f4$, $f5$, and $f6$. Once again, it is evident that the merger rate of PBHs is inversely proportional to their masses. In other words, the smaller the mass of PBHs participating in the merger event, the higher their number density per unit volume, and consequently, their total merger rate. As a result, the theoretical framework of $f(R)$ gravity imposes more stringent constraints on the fraction of PBHs if smaller masses of PBHs are considered. Additionally, it can be deduced from the figure that the field strength value $f_{R0}$ is also an actively contributing factor in constraining the fraction of PBHs. This establishes a direct relationship between the field strength and the stringency of the constraints on the fraction of PBHs. In this regard, the most stringent constraints can be obtained from $f(R)$ models of gravity with field strengths $f4$, $f5$, and $f6$, respectively. We have also quantified the overall results of our analysis for the merger rate of PBHs within the frameworks of $f(R)$ gravity and GR in Table \ref{table1}. The results show that the merger rate of PBHs, while considering the Einasto density profile, are on average about $60\%$ higher than the results obtained from the NFW density profile. Furthemore, it can be inferred that the most stringent constraint in this analysis is obtained from the models of $f(R)$ gravity, while considering the Einasto density profile and $M_{\rm PBH}\simeq 10\,M_{\odot}$, which, despite all theoretical uncertainties, enters the LIGO-Virgo-KAGRA sensitivity band for $f_{\rm PBH}\gtrsim 0.1$.
\section{Conclusions} \label{sec:v}
Primordial black holes are considered one of the most mysterious phenomena in astrophysics, and fundamental questions about their nature continue to be raised. As PBHs are expected to interact solely through gravity, and considering that a substantial collection of black holes demonstrates characteristics of perfect fluids on significantly large scales, PBHs present themselves as natural candidates for dark matter. On the other hand, due to their random distribution in the Universe, PBHs have the possibility of encountering each other, forming binaries, and eventually merging through the continuous propagation of GWs in the medium of dark matter halos. The dynamics of PBHs as dark matter candidates are expected to be influenced by the local and statistical properties of dark matter halos. However, a fundamental challenge arises as to whether dark matter halo models based on GR are good enough to accurately predict the merger rate of PBHs.

To address this question, in this study, we have calculated the merger rate of PBHs within the framework of $f(R)$ gravity and compared it with the corresponding results obtained from GR. To accomplish this task, we have initially established an appropriate framework for dark matter halo models that suits both GR and $f(R)$ gravity. We have introduced the definition of key parameters including the halo density profile, the halo concentration parameter, and the halo mass function. We have also discussed the field strength-dependent dynamical conditions induced by $f(R)$ gravity and introduced the appropriate density contrast parameter, $\delta_{\rm c}(z, M, f_{R0})$, and the mass function, $g^{f(R), \rm sx}(\sigma)$, for $f(R)$ gravity.

Furthermore, we have investigated the encounter conditions of randomly distributed PBHs within the context of dark matter halos. Under these assumptions, and considering $M_{\rm PBH}=30\,M_{\odot}$ and $f_{\rm PBH}=1$, we have calculated the merger rate of PBHs per halo while considering three models of $f(R)$ gravity, and compared the results with those obtained under GR. The results indicate that, in comparison to the outcome derived from GR, the merger rate of PBHs within each halo is higher for all the examined models under $f(R)$ gravity. This is because of the field strength-dependent dynamics of density fluctuations, $\delta_{\rm c}(z,M,f_{R0})$, which affects the formation and evolution of halo structures under $f(R)$ gravity. Additionally, we have witnessed a gradual reduction in the influence of field strenght-dependent dynamics, while moving from $f4$ to $f6$. 

Based on the PBH scenario and suitable halo mass functions, we have calculated the merger rate of PBHs per unit time and volume for three models of $f(R)$ gravity, and qualitatively and quantitatively compared them with the corresponding results obtained from GR. The results demonstrate that smaller-mass halos continue to exert a more significant influence on the merger rate of PBHs compared to larger halos. This phenomenon directly stems from the higher density of dark matter particles within subhalos, leading to a greater concentration within subhalos than in the host halos. Moreover, it can be deduced that across all the examined models under $f(R)$ gravity, the merger rate of PBHs surpasses what is obtained from GR. The direct connection between the merger rate of PBHs and the field strength is also evident.

The potential for binary PBH formation throughout the age of the Universe, stemming from their random distribution, serves as strong motivation to investigate the evolution of the PBH merger rate as a function of redshift. In light of this, we have specified the redshift-evolution of the PBH merger rate for three models of $f(R)$ gravity, and compared these findings with that obtained from GR. Consequently, the results demonstrate a direct correlation between the total merger rate of PBHs and the redshift in both models. In simpler terms, PBHs have exhibited a greater tendency to form binaries at higher redshifts compared to the present-time Universe. Furthermore, the direct correlation between the merger rate of PBHs and redshift becomes evident. This phenomenon can be elucidated by considering the influence of hierarchical dynamics of halo structures. The findings indicate that the redshift-dependent evolution of the PBH merger rate within $f(R)$ gravity models surpasses that obtained from the framework of GR. This suggests that, if we assume the validity of $f(R)$ gravity, the merger rate of PBHs will increase over time. Furthermore, it becomes evident that the influence of the field strength in enhancing the PBH merger rate will persist throughout the late-time Universe.

Lastly, we have computed the PBH merger rate for GR and three $f(R)$ gravity models as a function of their fraction with $M_{\rm PBH}=30\,M_{\odot}$, and then compared these results with the black hole mergers estimated by the LIGO-Virgo-KAGRA detectors during the latest observing run, i.e., $(17.9\mbox{-}44)\,{\rm Gpc^{-3}yr^{-1}}$. The results indicate that the merger rate of PBHs is inversely proportional to their masses, yet directly proportional to their fractions in both gravitational models. This phenomenon arises due to the inverse variation of PBH number density with their masses. Furthermore, it can be inferred that, when compared to the corresponding results derived from GR, the $f(R)$ gravity models satisfy the constraints imposed by GW data, particularly for lower values of the PBH fraction. Moreover, the significant impact of the field strength $f_{R0}$, originating from $f(R)$ gravity, in imposing stringent constraints on the PBH fraction becomes evident in our analysis. I have also provided a separate calculation for the merger rate of PBHs as a function of their mass and fraction in the context of $f(R)$ gravity. The findings indicate that, when the Einasto density profile is taken into account, the merger rate of PBHs is, on average, roughly $60\%$ higher than the findings from the NFW density profile. Furthermore, it can be deduced that the $f(R)$ gravity models, taking into account the Einasto density profile and $M_{\rm PBH} \simeq 10\,M_{\odot}$, yield the most stringent constraints. This mass range penetrates the LIGO-Virgo-KAGRA sensitivity region for $f_{\rm PBH} \gtrsim 0.1$ despite all theoretical uncertainties.

It should be noted that constraints on PBHs are subject to a variety of uncertainties, encompassing various gravitational frameworks, conditions that may have been imposed on the structures during collapse and formation, specific processes that can impact the growth or evaporation of PBHs (such as accretion and merger history), uncertainties arising from black hole formation scenarios, and their contribution to LIGO-Virgo-KAGRA mergers, as we have observed. While the presence of these factors might lead to computational errors, the development of future instruments and a deeper understanding of these unidentified processes may ultimately result in more stringent constraints on the fraction of PBHs.
\section*{ACKNOWLEDGEMENTS}
The author would like to gratefully acknowledge Marzieh Farhang from the Department of Physics of Shahid Beheshti University for constructive discussions and insightful comments. Also, the author thanks the Research Council of Shahid Beheshti University.

\bibliography{sample631}{}
\bibliographystyle{aasjournal}

\end{document}